\documentclass{article}
\usepackage{spconf,amsmath,graphicx}

\usepackage{amsmath,amssymb,amsfonts}
\usepackage{algorithm,algorithmic}
\usepackage{graphicx}
\usepackage{textcomp}
\usepackage{url}
\usepackage{color}
\usepackage{bm}

\usepackage{lipsum}
\usepackage{epstopdf}

\usepackage{amsopn}
\DeclareMathOperator{\diag}{diag}

\usepackage{amsthm}

\usepackage{enumerate}
\usepackage{cancel}
\usepackage{mathrsfs}
\usepackage{mathdots}
\usepackage{euscript}
\usepackage{amscd}
\usepackage{placeins}

\usepackage{multirow}
\usepackage{tikz}
\usetikzlibrary{arrows}
\usepackage{verbatim}

\usepackage{subcaption}

\newtheorem{theorem}{Theorem}

\theoremstyle{definition}

\newtheorem{example}{Example}

\theoremstyle{remark}
\newtheorem{remark}{Remark}

\newcommand{\bmat}{\left[ \begin{matrix}}
	\newcommand{\emat}{\end{matrix} \right]}

\DeclareMathOperator{\trace}{tr}

\DeclareMathOperator{\E}{{\mathbb E}}
\newcommand{\Rbb}{\mathbb R}

\newcommand{\Cbb}{\mathbb C}

\newcommand{\Zbb}{\mathbb Z}

\newcommand{\xb}{\mathbf  x}
\newcommand{\yb}{\mathbf  y}
\newcommand{\sbf}{\mathbf  s}  

\newcommand{\wb}{\mathbf  w}

\newcommand{\bb}{\mathbf  b}

\newcommand{\ub}{\mathbf  u}

\newcommand{\zerob}{\mathbf 0}

\newcommand{\thetab}{\boldsymbol{\theta}}

\DeclareMathOperator{\range}{Range}
\DeclareMathOperator{\rank}{rank}

\newcommand{\Acal}{\mathcal{A}}

\newcommand{\Ical}{\mathcal{I}}
\newcommand{\Hcal}{\mathcal{H}}

\renewcommand{\d}{\mathrm{d}}

\newcommand{\SNR}{\mathrm{SNR}}

\newcommand{\srm}{\mathrm{s}}

\title{When atomic norm meets the G-filter: A general framework for line spectral estimation}
\name{Bin Zhu and Jiale Tang
\thanks{This work was support supported in part by Shenzhen Science and Technology Program (Grant No.~202206193000001-20220817184157001), and the ‘‘Hundred-Talent Program’’ of Sun Yat-sen University.}}
\address{School of Intelligent Systems Engineering, Sun Yat-sen University, 518107 Shenzhen, China.}
\begin{document}
%
\maketitle

\begin{abstract}
	
This paper proposes a novel approach for line spectral estimation which combines Georgiou's filter bank (G-filter) with atomic norm minimization (ANM).
A key ingredient is a Carath\'{e}odory--Fej\'{e}r-type decomposition for the covariance matrix of the filter output. 
The resulting optimization problem can be characterized via semidefinite programming and contains the standard ANM for line spectral estimation as a special case.
Simulations show that our approach outperforms the standard ANM in terms of recovering the number of spectral lines when the signal-to-noise ratio is no lower than 0 dB and the G-filter is suitably designed.

\end{abstract}

\section{Introduction}



It is well known that the spectrum of a sinusoidal signal consists of spectral lines (Dirac impulses).
The problem of line spectral estimation concerns reconstruction of the spectral lines from a finite number of signal measurements \cite{stoica2005spectral}.
In the time domain, it is equivalent to estimating the amplitude and frequency of each component in the sinusoidal signal.
Such a problem is of fundamental importance in signal processing with numerous applications notably in radars and sonars where the problem is also called ``direction-of-arrival estimation'' or ``array processing'' \cite{van2004optimum}.


Besides classic FFT-based methods and subspace methods, one of the mainstream approaches nowadays is known as \emph{atomic norm minimization} (abbreviated as ANM) which is inspired by ideas from \emph{compressed sensing}, see e.g., \cite{tang2013compressed,bhaskar2013atomic,Zhu-M2-LineSpec}.
Indeed, the sinusoids can be viewed as a \emph{spectrally sparse} signal and the atomic norm can be used to promote such sparsity. 
The success of ANM for frequency estimation is guaranteed by a mathematical result called \emph{Carath\'{e}odory--Fej\'{e}r} (abbreviated as C--F) decomposition
for positive semidefinite Toeplitz matrices, cf. e.g., \cite{Grenander_Szego}. 
Exploiting the C--F decomposition, the unknown frequencies can be encoded in the Toeplitz covariance matrix which ultimately convexifies the optimization problem.


In \cite{georgiou2000signal,amini2006tunable}, Georgiou provided a substantial generalization of the C--F decomposition from Toeplitz matrices to output covariance matrices corresponding to a class of stable linear filter banks which we call ``G-filter''. 
It then seems natural to incorporate this generalized decomposition into the ANM framework for frequency estimation.
Indeed, we show in this paper  that such a combination can be achieved,
and the G-filter version of the ANM approach formally generalizes the standard ANM.
Moreover, simulations indicate that our generalized approach performs better than the standard ANM when the signal-to-noise ratio (SNR) is not too low and the G-filter selects a desired frequency band.


The rest of this paper is organized as follows. The frequency estimation problem is reviewed in Sec.~\ref{sec:prob}. Georgiou's filter bank and a general signal model are described in Sec.~\ref{sec:G-filter}. 
The C--F-type decomposition for output covariance matrices of G-filters are discussed in Sec.~\ref{sec:C-F-type_decomp}.
The ANM problem with a G-filter integrated for frequency estimation is treated in Sec.~\ref{sec:ANM}.
Extensive numerical simulations are provided in Sec.~\ref{sec:sims}.
Finally, Sec.~\ref{sec:conc} concludes the paper.

%




\section{Frequency estimation problem}\label{sec:prob}

Suppose that we have measured some complex sinusoids (cisoids) in noise:
\begin{equation}\label{signal_model}
y(t) = s(t) + w(t) = \sum_{k=1}^m a_k \, e^{i \theta_k t} + w(t),
\end{equation}
where, $t = 0, 1, \dots, L-1$, $s$ is a linear combination of $m$ complex exponentials $\{e^{i \theta_k t}\}$ with unknown angular frequencies $\{\theta_k\}\subset\Ical:=[0, 2\pi)$, and $w$ is the additive noise. 
The coefficients in $\{a_k\}$ are complex amplitudes.
The signal model \eqref{signal_model} can also be put in a vector form $\yb=\sbf+\wb$ where
\begin{equation}\label{y_meas_vec}
\sbf := \bmat s(0) & \cdots & s(L-1) \emat^\top = \sum_{k=1}^m G_0(e^{i\theta_k}) a_k,
\end{equation}
and the components of $\yb$ and $\wb$ are collected in the same order as $\sbf$.
The vector-valued function 
\begin{equation}\label{map_array_geom}
G_0(e^{i\theta}) := \bmat 1 & e^{i\theta} & \cdots & e^{i(L-1)\theta}\emat^\top
\end{equation}
is determined by a uniform linear array for the measurement.


The standard frequency estimation problem consists of determining the number $m$ of unknown frequencies, and constructing an estimate of $\{\theta_k\}$ from the finite measurements $y$ in \eqref{signal_model}.
When the frequencies are obtained, the amplitudes $\{a_k\}$ can be estimated via least squares. 
Therefore, we are mostly interested in frequency estimation.

\section{General signal model via G-filtering}\label{sec:G-filter}



The G-filter \cite{georgiou2000signal} is defined by the equation
\begin{equation}\label{filter_bank}
\xb(t+1) = A\xb(t) + \bb y(t), \quad t\in\Zbb,
\end{equation}
where $y$ is a scalar input and $\xb$ is a vector output of size $n$. The matrix $A\in\Cbb^{n\times n}$ has a spectral radius $\rho(A)<1$, the vector $\bb\in\Cbb^n$, and $(A, \bb)$ is a \emph{reachable} pair, namely
$\rank \bmat \bb & A\bb & \cdots & A^{n-1}\bb\emat = n$.
The transfer function of the filter \eqref{filter_bank} is
\begin{equation}\label{trans_func_filter_bank}
G(z) = \bmat g_1(z) & \cdots & g_n(z) \emat^\top = (zI-A)^{-1} \bb 
\end{equation}
where $z$ can be interpreted as a shift operator $\xb(t) \mapsto \xb(t+1)$.

\begin{example}
	\label{ex_delay_filt_bank}
	Take $A$ as the $n\times n$ Jordan block with a complex constant $p$ on the main diagonal such that $|p|<1$:
	\begin{equation}\label{filt_paras_Jordan}
	A =\bmat p & 1 & \cdots & 0 \\
	\vdots & \ddots & \ddots & \vdots \\
	0 & \cdots &p & 1 \\
	0 & \cdots & 0 & p\emat,
	\ \text{and}\
	\bb =\bmat 0 \\ \vdots \\ 0 \\ 1 \emat.
	\end{equation} 
	If $p=0$, we have the components of $G(z)$ in \eqref{trans_func_filter_bank}
	\begin{equation}\label{delay_filter}
	g_k(z) = z^{-n-1+k}, \quad k=1, 2, \dots, n,
	\end{equation}
	which defines a \emph{delay} filter bank.
	When the filter size $n$ is equal to the signal length $L$ in \eqref{signal_model}, the delay filter bank results in $G_0(e^{i\theta})$ in \eqref{map_array_geom} up to a scaling factor $e^{-in\theta}$.
\end{example}

\begin{figure}[h]
	\centering
	\tikzstyle{int}=[draw, minimum size=2em]
	\tikzstyle{init} = [pin edge={to-,thin,black}]
	\begin{tikzpicture}[node distance=2cm,auto,>=latex']
	\node [int] (a) {$\ G(z)\ $};
	\node (b) [left of=a, coordinate] {};
	\node (c) [right of=a] {};
	\path[->] (b) edge node {$y(t)$} (a);
	\path[->] (a) edge node {$\xb(t)$} (c);
	\end{tikzpicture}
	\caption{A general signal model.}
	\label{fig:model_general}
\end{figure}
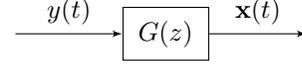

The above example indicates that we can replace the signal model \eqref{y_meas_vec} with a more general form involving the G-filter as illustrated by Fig.~\ref{fig:model_general}.
Indeed, the output of the filter is related to the input via the following expression:
\begin{equation}\label{state_plus_noise}
\begin{aligned}
\xb(t) & = G(z) [s(t)+w(t)] \\
& := \int_{\Ical} G(e^{i\theta}) e^{i\theta t} \left[ \d\hat{s}(\theta) + \d\hat{w}(\theta) \right] = \tilde{\sbf}(t) + \tilde{\wb}(t)
\end{aligned}
\end{equation}
where, $\d\hat{s}(\theta) = \sum_{k=1}^m a_k\delta(\theta-\theta_k)\d\theta$ and $\d\hat{w}(\theta)$ are the \emph{spectral measures} of the cisoidal signal $s(t)$ and the noise $w(t)$, respectively,
\begin{equation}\label{state_atomic_decomp}
\tilde{\sbf}(t):=\sum_{k=1}^m G(e^{i\theta_k}) c_k(t)
\end{equation}
is the filtered cisoids (similar to \eqref{y_meas_vec}) with amplitudes
$\{c_k(t) = a_k e^{i\theta_k t}\}$,
and $\tilde{\wb}(t)$ is the filtered noise vector. If $\Phi_w(\theta)$ is the power spectral density of the input noise $w(t)$, then $\tilde{\wb}$ has a matricial spectral density 
$G(e^{i\theta}) \Phi_w(\theta) G^*(e^{i\theta})$.
Here $G^*(z):=\bb^*(z^{-1}I-A^*)^{-1}$ is the conjugate of 
$G(z)$ 
in \eqref{trans_func_filter_bank}, and consequently, we have $\left[G(e^{i\theta})\right]^*=G^*(e^{i\theta})$. 

In the remaining part of this paper, we shall make the innocuous assumption that the number of cisoids (spectral lines) in the signal is less than the size of the G-filter, namely $m<n$. Such an inequality can be easily achieved by increasing the filter size $n$.


\begin{remark}[Filtering a finite-length signal]\label{rem_input_filtering}
	
		
	Given a cisoidal signal $y(t)$ of the form \eqref{signal_model}, we can carry out the filtering operation as described by \eqref{filter_bank} in a straightforward manner with an initial condition $\xb(0)=\zerob$. In order to remedy the transient effect of the initial condition, we simply discard the first $L_{\srm}$ filtered samples such that $\|A^{L_{\srm}}\|<\varepsilon$ with a predefined threshold $\varepsilon>0$, as suggested in \cite[p.~2664]{amini2006tunable}.
	In addition, we assume that only \emph{one} output vector is available after the truncation which is in line with the standard setup of a delay filter bank. 

\end{remark}


\section{Carath\'{e}odory--Fej\'{e}r-type decomposition}\label{sec:C-F-type_decomp}

In view of Fig.~\ref{fig:model_general}, the output covariance matrix is defined as $\Sigma:=\E \left[\xb(t)\xb(t)^*\right]$
which admits an integral representation 
\begin{equation}\label{state_cov_mat}
	\Sigma = \frac{1}{2\pi} \int_{\Ical} G(e^{i\theta})\, \d\mu_y(\theta)\, G^*(e^{i\theta})
\end{equation}
because of the filtering operation.
In \eqref{state_cov_mat}, the nonnegative measure $\d\mu_y$ represents the power spectrum of the input $y$.
Spectral estimation based on the output covariance matrix has been studied intensively since the beginning of this century, see e.g., \cite{georgiou2000signal,BGL-THREE-00,amini2006tunable}. 
Next we briefly review the algebraic structure of an output covariance matrix dictated by the G-filter and a related decomposition theorem since they will be fundamental for later development.

Let $\Gamma$ be a linear operator
that sends a \emph{signed} measure $\d\mu$ to a matrix $({1}/{2\pi}) \int_{\Ical} G(e^{i\theta})\, \d\mu(\theta)\, G^*(e^{i\theta})\in\Hcal_n$ where $\Hcal_n$ represents the linear space (over the reals) of Hermitian matrices of size $n$.
Then a consequence of \eqref{state_cov_mat} is $\Sigma\in\range\Gamma$, and the latter is a linear subspace of $\Hcal_n$. 
The set membership $\Sigma\in\range\Gamma$ characterizes the feasibility of 
optimization problems to be formulated in the next section. 
An equivalent characterization \cite[Prop.~3.2]{ferrante2012maximum} 
is given by the equality
\begin{equation}\label{equal_constraint}
	(I-\Pi_{\bb})(\Sigma-A\Sigma A^*)(I-\Pi_{\bb}) = O,
\end{equation}
where $\Pi_{\bb}:=\bb\bb^*/(\bb^*\bb)$ is a projection matrix and $(A, \bb)$ is the G-filter parameter in \eqref{filter_bank}.
It turns out that \eqref{equal_constraint} is more convenient for algorithmic implementation.

	We remark that $\range\Gamma$ reduces to the set of all Hermitian Toeplitz matrices in the case of Example~\ref{ex_delay_filt_bank} with $p=0$.
	It is well known that a positive semidefinite Toeplitz matrix can be decomposed \emph{\`{a} la} Carath\'{e}odory and Fej\'{e}r \cite{Grenander_Szego}.
	Such a decomposition of Carath\'{e}odory--Fej\'{e}r (C--F) type has been generalized to output covariance matrices in \cite{georgiou2000signal}, and result is recalled as follows.
	For convenience, a positive semidefinite matrix $A$ is written as $A\geq 0$.

\begin{theorem}[C--F-type decomposition \cite{georgiou2000signal}]\label{thm:C-F_type_decomp}
	Let $\Sigma\geq 0$ be an output covariance matrix in the sense of \eqref{state_cov_mat} having rank $r< n$. Then it admits a unique decomposition of the form
	\begin{equation}\label{C-F-type-decomp}
	\Sigma = \sum_{k=1}^{r} \rho_k G(e^{i\theta_k}) G^*(e^{i\theta_k})
	\end{equation}
	where each $\rho_k>0$, and the frequencies $\{\theta_k\in\Ical\}$ are distinct for $k=1, 2,\dots, r$. 
\end{theorem}

The above C--F-type decomposition is numerically computable, see \cite[Prop.~2]{georgiou2000signal}. Some computational steps are provided as follows in which we are only interested in the frequencies $\{\theta_k\}$. First, we compute the spectral decomposition of $\Sigma=U \diag\{\lambda_1, \cdots, \lambda_r, 0, \dots, 0\} U^*$
where $U$ is unitary and $\lambda_k>0$ for $k=1, \dots, r$. Let $\ub_k$ be the $k$-th column of $U$, and eigenvector matrix is partitioned as 
$U = \bmat U_{1:r} & U_{r+1:n}\emat$
where the symbol $U_{k:\ell}$ with $k\leq \ell$ denotes a matrix whose columns are $\ub_k, \ub_{k+1}, \dots, \ub_\ell$.
Then, we construct a rational function which is nonnegative on the unit circle:
\begin{equation}\label{rat_func_sym}
d(z, z^{-1}) = G^*(z) U_{r+1:n}\, U_{r+1:n}^* G(z).
\end{equation}
The parameters $\{\theta_k\}$ in \eqref{C-F-type-decomp} correspond to the distinct roots of $d(z, z^{-1})$ on the unit circle, i.e., of the form $e^{i\theta_k}$, and there are exactly $r$ such roots.

\section{Atomic norm minimization approach}\label{sec:ANM}

To simplify the presentation, we first assume that the signal model \eqref{signal_model} is noiseless, i.e., $w(t)\equiv0$.
Then the filtered signal is precisely \eqref{state_atomic_decomp}
which is a linear combination of certain elements, called ``atoms'', from the atomic set
\begin{equation}\label{dictionary}
\Acal := \{G(e^{i\theta}) : \theta\in \Ical\}.
\end{equation}
More precisely, the right-hand side of \eqref{state_atomic_decomp} is called an \emph{atomic decomposition} which contains the unknown frequency $\theta_k$
in the atom $G(e^{i\theta_k})$. In view of Remark~\ref{rem_input_filtering}, it is assumed that we only have access to the output $\xb(t)\equiv\tilde \sbf(t)$ at one single time instance $t$, and we simply write $\tilde{\sbf}=\tilde\sbf(t)$.
The \emph{atomic norm} of a noiseless measurement vector is defined as
\begin{equation}\label{atomic_norm}
\|\tilde\sbf\|_{\Acal} := \inf_{\substack{c_k\neq 0,\\ \theta_k\in\Ical}} \left\{ \sum_{k} |c_k| \|G(e^{i\theta_k})\| : \tilde \sbf = \sum_{k} G(e^{i\theta_k}) c_k \right\}
\end{equation}
which can be interpreted as the spectral version of a weighted $\ell_1$ norm. 
Hence it can promote sparsity in the frequency domain in the sense that the number of selected atoms 
should be as few as possible.
In addition, the atomic norm is very flexible because the atoms are parametrized by $\theta$ in a continuum.

The next result, whose proof can be constructed along the lines of \cite[Prop.~II.1]{tang2013compressed}, 
shows how to compute the atomic norm $\|\tilde\sbf\|_{\Acal}$ via \emph{semidefinite programming} (SDP).
The latter can be handled with standard convex optimization tools \cite{bv_cvxbook}.


\begin{theorem}\label{thm_AN_noiseless}
	Given one output vector $\tilde\sbf \equiv \tilde\sbf(t)$ of the G-filter whose input is some noiseless cisoids, the atomic norm $\|\tilde\sbf\|_\Acal$ is equal to the optimal value of the semidefinite program
	\begin{subequations}\label{AN_semidef_program}
		\begin{align}
		& \underset{\substack{\tau\in\Rbb,\ \Sigma\in\Hcal_n}}{\text{minimize}}
		& & \frac{1}{2} (\tau + \trace\Sigma) \label{obj_noiseless} \\
		& \text{subject to}
		& & \bmat \tau&\tilde\sbf^* \\ \tilde\sbf &\Sigma \emat \geq 0, \label{LMI_constraint} \\
		& & &  \quad \text{and}\quad \eqref{equal_constraint}.
		\end{align}
	\end{subequations}
\end{theorem}


The general case with noise can be treated in the style of \cite{bhaskar2013atomic}. To this end, 
we set up a \emph{regularized} optimization problem of minimizing $\frac{1}{2} \|\xb-\tilde\sbf\|^2 + 
\lambda \|\tilde\sbf\|_{\Acal}$ over $\tilde\sbf \in \Cbb^n$,
where $\xb$ is the noisy measurement vector and $\lambda>0$ is a regularization parameter. The above problem also admits a SDP formulation:
\begin{subequations}\label{noisy_SDP}
	\begin{align}
	& \underset{\substack{\tau\in\Rbb,\ \tilde\sbf \in \Cbb^n \\  \Sigma\in\Hcal_n}}{\text{minimize}}
	& & \frac{1}{2} \|\xb-\tilde\sbf\|^2 + \lambda (\tau + \trace\Sigma) \label{obj_noisy} \\
	& \text{subject to}
	& & \eqref{LMI_constraint}  \quad \text{and}\quad \eqref{equal_constraint}.
	\end{align}
\end{subequations}
After solving the SDP, 
the frequency estimates $\{\hat{\theta}_k\}$ are computed from the C--F-type decomposition of the optimal $\hat\Sigma$ in the sense of Theorem~\ref{thm:C-F_type_decomp}.

%

\section{Simulations}\label{sec:sims}

In this section, we perform numerical simulations for our frequency estimation approach in comparison with the standard ANM.
Some details of implementation are provided next.

\noindent{\bf Construction of a G-filter.}
We use the G-filter in Example~\ref{ex_delay_filt_bank} with one repeated pole at $p=\rho e^{i\varphi}$ of multiplicity $n$, and impose the normalization condition
$A A^* + \bb \bb^* = I$
following the procedure in \cite[Sec.~VII-F]{amini2006tunable}.
Such a filter bank should select a frequency band $[\theta_{\ell}, \theta_{u}]$ which represents our \emph{a priori} knowledge about the locations of the cisoids. By band selection, we mean that the filter ``gain'' $\|G(e^{i\theta})\|$ is relatively large inside the band.
The parameters $(\rho, \varphi)$ are determined as per \cite[p.~2667]{amini2006tunable}.

\noindent{\bf Filtering a finite-length signal $y(t)$.}
With reference to Remark~\ref{rem_input_filtering}, we set the threshold $\varepsilon=10^{-3}$ to determine the number $L_\srm$ of discarded filter outputs.
Moreover, we assume that the signal length $L$ of $y$ is small such that $L-L_{\srm}=1$. 


\noindent{\bf Choice of the regularization parameter $\lambda$ in the noisy case.} 
According to \cite{bhaskar2013atomic}, we must take $\lambda\geq \E\|\tilde\wb\|_{\Acal}^*$ in order to recover the signal vector $\tilde\sbf$ in a stable manner.
In the standard ANM setting, an estimate for $\E\|\tilde\wb\|_{\Acal}^*$ can be explicitly computed using \emph{Bernstein's inequality}, see \cite{bhaskar2013atomic}. However, in the general case with the atomic set \eqref{dictionary}, it seems that new techniques must be developed for the estimation of $\E\|\tilde\wb\|_{\Acal}^*$.
In what follows, we use the \emph{heuristic} value
$\lambda =\frac{\sigma}{2} \sqrt{n\log n}$
which is the dominant term of the regularization parameter in the case of a delay filter bank.
It still remains to estimate the noise variance $\sigma^2$ for which we adopt the procedure in \cite{bhaskar2013atomic}.
%

\noindent{\bf Solving the SDPs.}
The SDP \eqref{noisy_SDP} is solved using CVX, a package for specifying and solving convex programs \cite{cvx,gb08}. 
The frequency estimate $\hat{\thetab}=(\hat\theta_1,\dots,\hat\theta_{\hat{r}})$ is computed from the optimal $\hat{\Sigma}$ via the C--F-type decomposition in Sec.~\ref{sec:C-F-type_decomp}
where $\hat r$ is the \emph{numerical rank} of $\hat{\Sigma}$.
Therefore, then number $\hat{r}$ is just our estimate of $m$, the number of cisoids in $y(t)$.
In fact, the numerical rank is computed as follows.
Let $\hat\lambda_1\geq \hat\lambda_2 \geq \cdots \geq \hat\lambda_n$ be the eigenvalues of $\hat{\Sigma}$. Then $\hat r$ is equal to the first positive integer $k$ such that $\hat\lambda_{k+1}<10^{-3}$ or $\hat\lambda_k/\hat\lambda_{k+1}>10^3$.

%

\noindent{\bf Simulation results.} 
	The number of cisoids in the signal $y$ is $m=3$, and the signal length is $L=98$. The true frequencies are set as $\theta_1=\theta_0-2(2\pi/L)$, $\theta_2=\theta_0$, and $\theta_3=\theta_0+2(2\pi/L)$ where $\theta_0\in\{1.5, 1.6, \dots, 2.5\}$. The amplitudes are $a_k=1e^{i\varphi_k}$, $k=1,2,3$ where $\varphi_k$'s are uniform random variables
	in $[0, 2\pi]$. 
	The frequencies are separated at a distance equal to twice of the resolution limit $2\pi/L$ of the FFT method, which proves to create difficulties for the standard ANM, as revealed by our results.
	The SNR is defined as $10\log_{10}(1^2/\sigma^2)$ dB. The candidate values for the SNR are $-3, 0, 3, 6, 9$ dB. The noise variance $\sigma^2$ is determined once the SNR is fixed, and is then used to generate complex Gaussian white noise $w(t)$.
	
\begin{figure}
		\centering
		\includegraphics[width=0.11\textwidth]{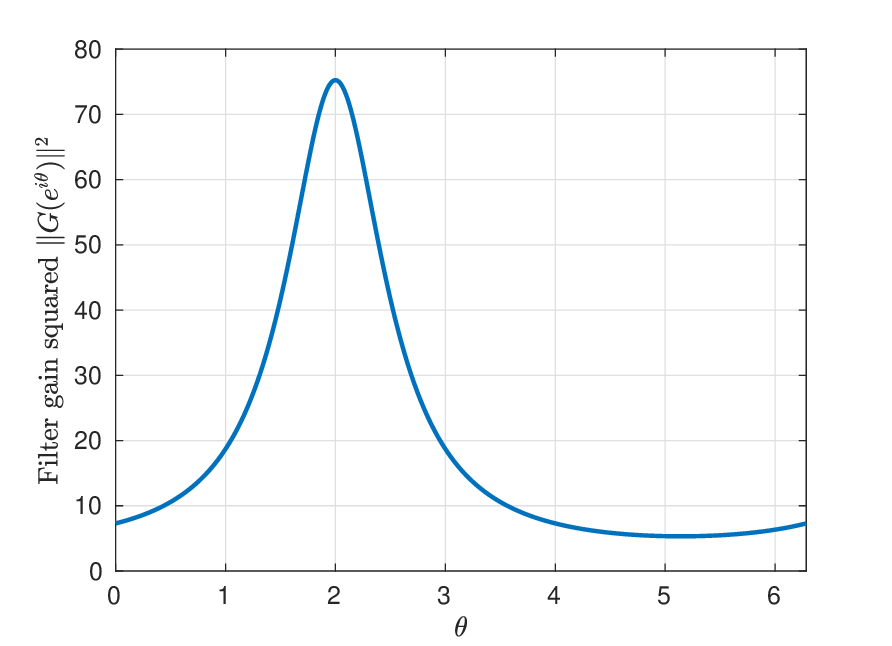}
		\includegraphics[width=0.11\textwidth]{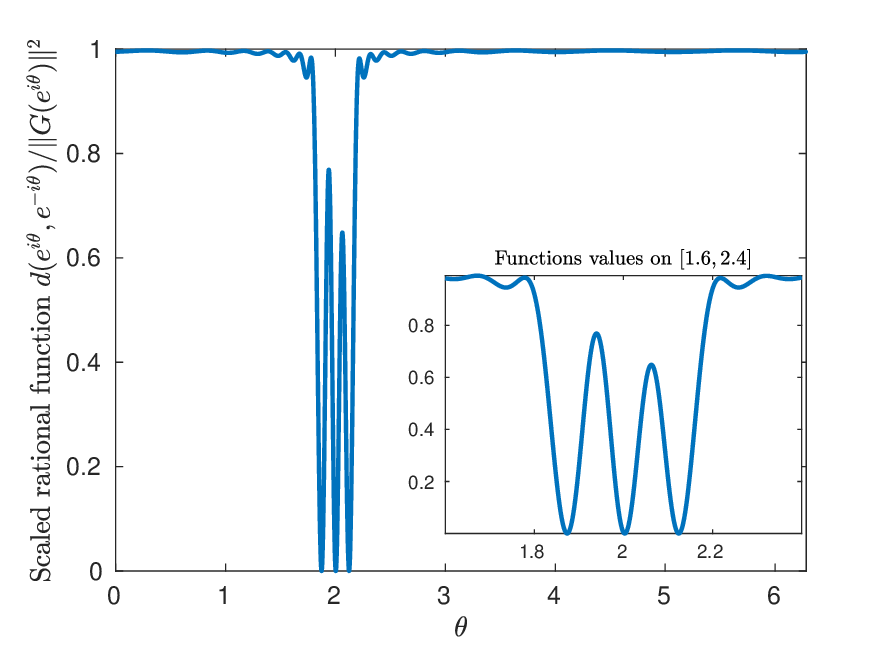}
\includegraphics[width=0.11\textwidth]{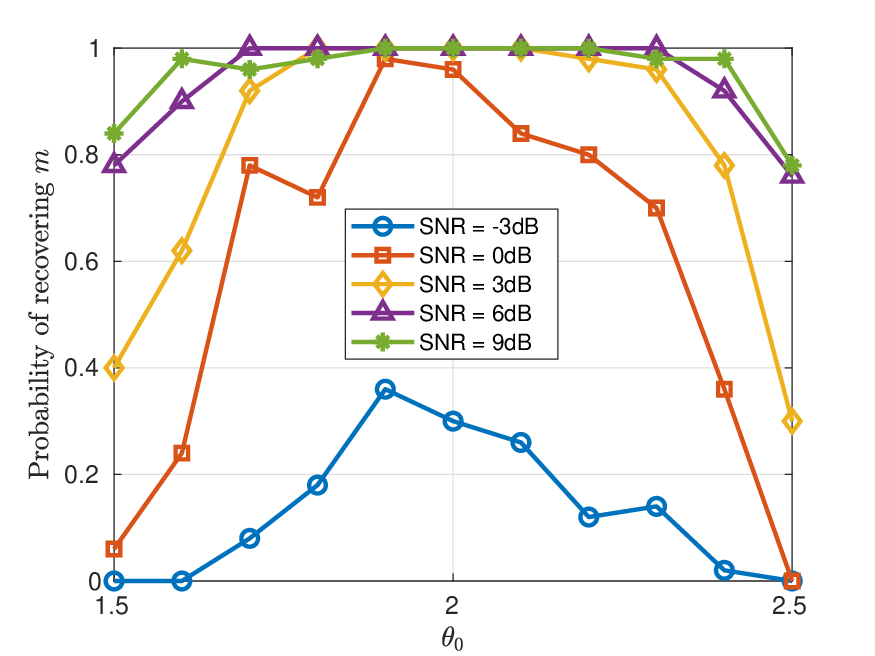}
\includegraphics[width=0.11\textwidth]{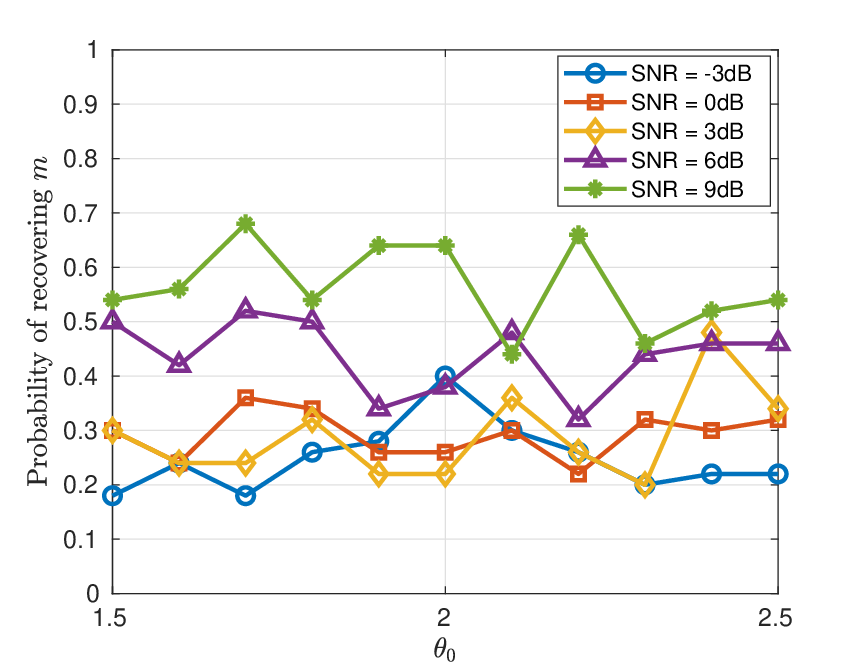}
	
	\caption{\emph{From left to right}: Squared gain $\|G(e^{i\theta})\|^2$ of the G-filter, graph of the scaled rational function $\bar d(e^{i\theta},e^{-i\theta})$,
	probability of successfully recovering the number of cisoids with our approach, and similar probability with the standard ANM, respectively.
	}
	\label{fig:graphs_funcs}
\end{figure}

	We 
	use a G-filter of size $n=20$ with a repeated pole at $p=0.58e^{i2}$ which selects the frequency band $\Ical_1=[1.75, 2.25]$, see \cite[Ex.~2]{amini2006tunable}.
	The graph of $\|G(e^{i\theta})\|^2$
	is shown in the first panel of Fig.~\ref{fig:graphs_funcs}. Clearly, the curve is unimodal in $\Ical$, and has a peak at $\theta=2$.
	The number of truncated output vectors of the filter is $L_{\srm}=L-1=97$. 
	
%

\begin{figure}
			\includegraphics[width=.2\textwidth]{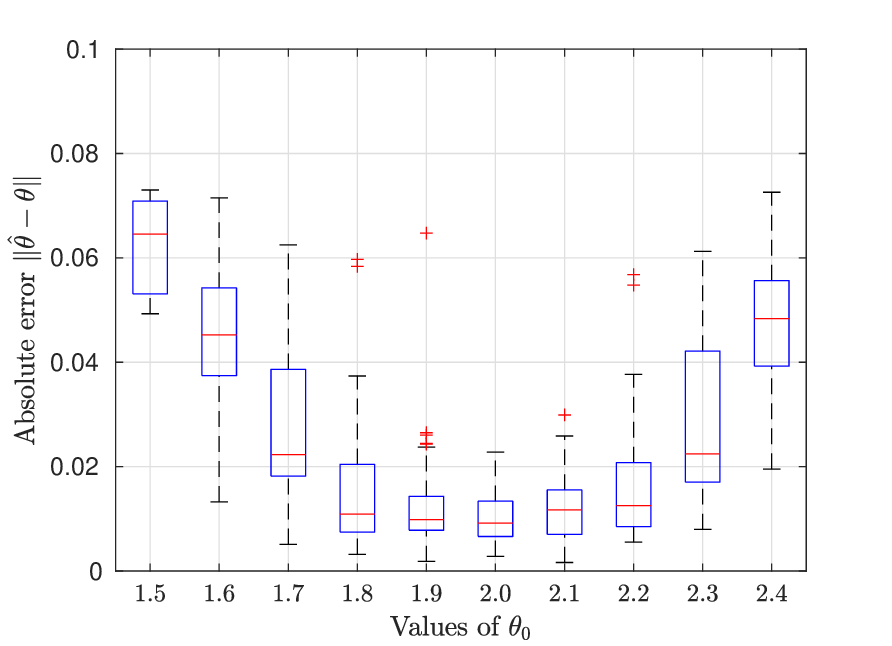}
		\hfill
			\includegraphics[width=.2\textwidth]{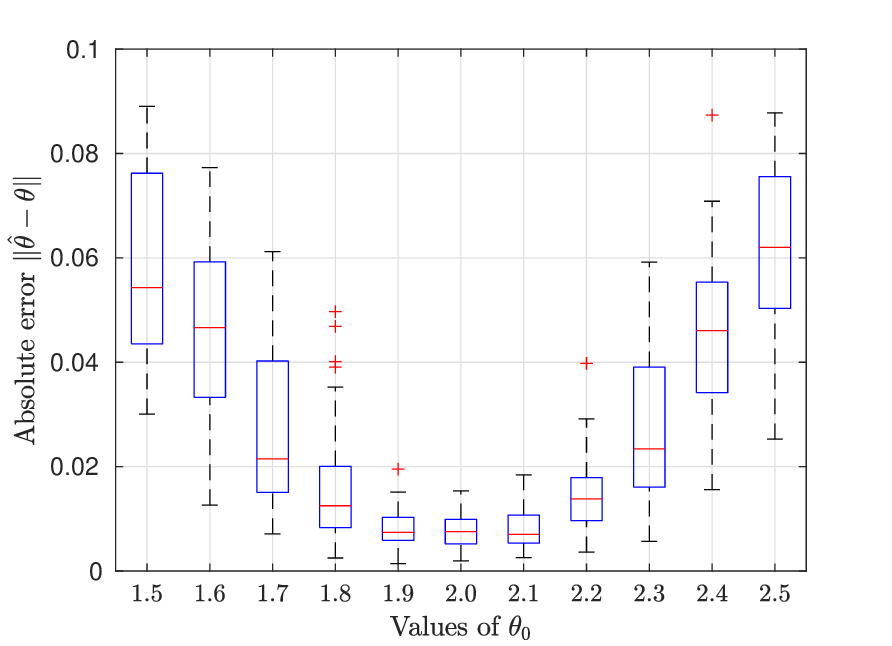}
		\\
			\includegraphics[width=.2\textwidth]{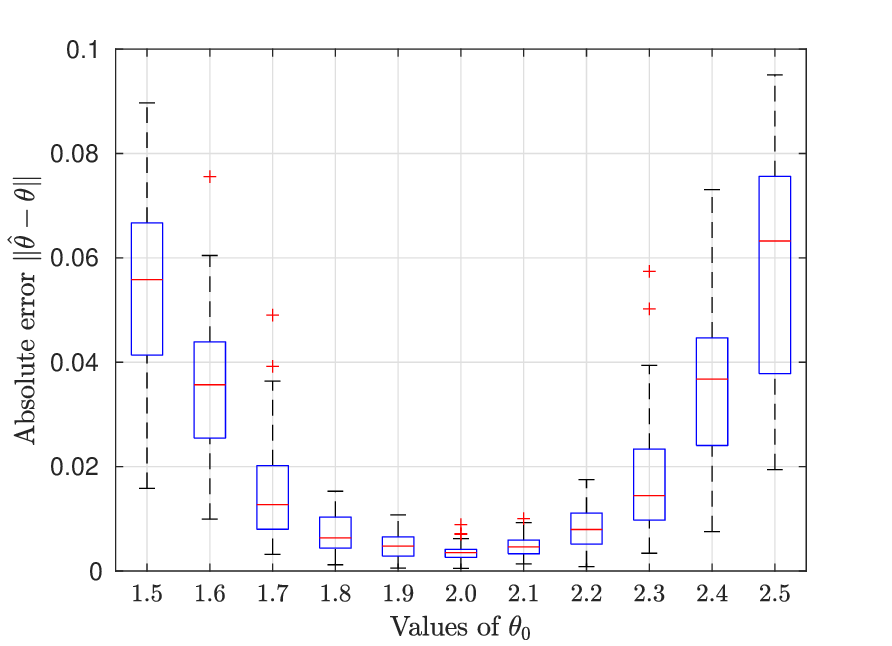}
		\hfill
			\includegraphics[width=.2\textwidth]{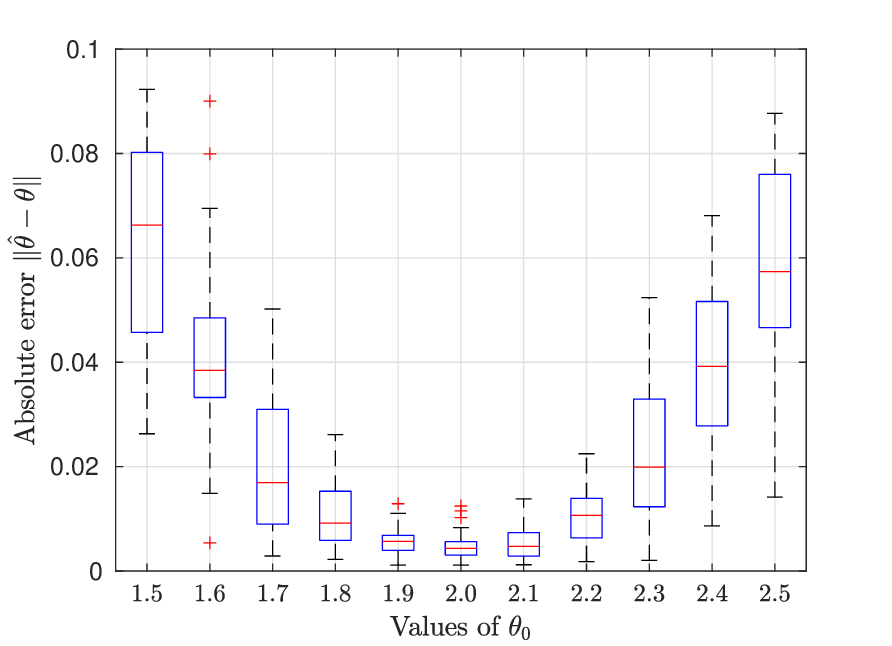}
		
		\caption{\emph{Clockwise}: Absolute errors $\|\hat{\thetab}-\thetab\|$ of frequency estimation in each Monte Carlo simulation under $\SNR = 0, 3, 6$, and $9$ dB, respectively.
			 Notice that the box for $\theta_0=2.5$ is not displayed in the upperleft panel because it is empty, see the third panel of Fig.~\ref{fig:graphs_funcs}. 
			 Errors for $\SNR=-3$ dB are not shown for a similar reason.}
		\label{fig:sim_results_three_freqs}
	\end{figure}

	For each value of $\theta_0$, we run a Monte Carlo simulation which contains $50$ repeated trials to evaluate the performance of our approach. In the two panels on the right of Fig.~\ref{fig:graphs_funcs}, we show the ``probability'' of successfully recovering the number of cisoids which is defined as $\#\{\text{trials} : \hat{r}=m=3\}/50$ where $\#\{\cdot\}$ denotes the cardinality of a set.
	 We observe that the probability is close to $1$ when $1.8\leq \theta_0\leq 2.2$ and $\SNR\geq 3$ dB.
	 In contrast, the recovery probability of the standard ANM is below $0.7$ even when the $\SNR = 9$ dB.
	 In order to illustrate the frequency extraction procedure as described at the end of Sec.~\ref{sec:C-F-type_decomp}, we show in the second panel of Fig.~\ref{fig:graphs_funcs} the scaled rational function $\bar{d}(e^{i\theta},e^{-i\theta}) := d(e^{i\theta},e^{-i\theta})/\|G(e^{i\theta})\|^2$ which is computed from the optimal $\hat{\Sigma}$ in one trial with $\theta_0=2$. 
	 The points of minimum of $\bar{d}$ are taken as estimates of the frequencies . 
	 In addition, the absolute error of frequency estimation $\|\hat{\thetab}-\thetab\|$ is computed for all successful trials where $\thetab=(\theta_1, \theta_2, \theta_3)$ is the true frequency vector and $\hat{\thetab}$ the estimate. These errors are depicted in Fig.~\ref{fig:sim_results_three_freqs} using the $\mathtt{boxplot}$. A general trend is that the errors decrease as the SNR increases and the smallest errors center around $\theta_0=2$ which is the midpoint of the selected band $\Ical_1$.

%
%
	

\section{Conclusion}\label{sec:conc}

This paper integrates Georgiou's filter bank (G-filter) into the atomic norm minimization (ANM) framework for frequency estimation. 
A Carath\'{e}odory--Fej\'{e}r-type decomposition is used to encode frequencies into the output covariance matrix, which results in convex semidefinite programs.
Simulations show that the G-filter version of ANM outperforms the standard ANM in terms of detecting the correct number of cisoids.
The errors of frequency estimation are small provided that the SNR is not too low, and the true frequencies
are close to the center of the band selected by the G-filter.



\bibliographystyle{IEEEbib-abbrev}
\bibliography{refs}

\end{document}